# Drone-based all-weather entanglement distribution

Hua-Ying Liu*, Xiao-Hui Tian*, Changsheng Gu*, Pengfei Fan*, Xin Ni, Ran Yang, Ji-Ning Zhang, Mingzhe Hu, Yang Niu, Xun Cao, Xiaopeng Hu, Gang Zhao, Yan-Qing Lu, Zhenda Xie[+], Yan-Xiao Gong[+], and Shi-Ning Zhu[+]

*National Laboratory of Solid State Microstructures, School of Electronic Science and Engineering, School of Physics, College of Engineering and Applied Sciences, and Collaborative Innovation Center of Advanced Microstructures, Nanjing University, Nanjing 210093, China*

\* These authors contributed equally to this work.

+email:

xiezhenda@nju.edu.cn

gongyanxiao@nju.edu.cn

zhusn@nju.edu.cn

The quantum satellite is a cornerstone towards practical free-space quantum network and overcomes the photon loss over large distance. However, challenges still exist including real-time all-location coverage and multi-node construction, which may be complemented by the diversity of modern drones. Here we demonstrate the first drone-based entanglement distribution at all-weather conditions over 200 meters (test field limited), and the Clauser-Horne-Shimony-Holt S-parameter exceeds 2.49 $\pm$ 0.09, within 35 kg take-off weight. With symmetric transmitter and receiver beam apertures and single-mode-fiber-coupling technology, such progress is ready for future quantum network with multi-node expansion. This network can be further integrated in picture-drone sizes for plug-and-play local-area coverage, or loaded onto high-altitude drones for wide-area coverage, which adds flexibility while connecting to the existing satellites and ground fiber-based quantum network.

Quantum network is the ultimate solution for secure data transferring, where an unknown quantum state cannot be copied or amplified[1]. However, this quantum nature sets a poor tolerance to photon loss in the network channels, which limits the communication distance and data rate with the existing fiber networks[2-4]. Therefore, the free-space quantum network [5-7] is a natural solution to overcome this problem. So far, efforts have been focused on the satellite-based free-space quantum networks[8-11], which enable large-distance satellite-ground links, taking advantage of the low scattering loss in the empty space. Based on comprehensive ground and aerial-based experimental tests[12-17], up to 1203 km entanglement distribution has been achieved with such quantum satellites[9, 18-21]. However, the existing quantum satellites may not fulfill all the requirements of a practical quantum network. First, the low-orbital satellites can only establish the quantum data link for certain ground locations within a limited time window. Second, it is difficult and expensive to build a scalable quantum network with multi nodes using only satellites, as the link distance between the satellites is limited by the spaceborne beam apertures. On the other hand, the drone, or unmanned aerial vehicle (UAV)[22, 23] has undergone an explosive development[24, 25] because of the breakthrough in the automatic flight control system and artificial intelligence. It covers a take-off weight from a few grams to tens of tons, a cruising altitude from meters to over 20 km, and a flight duration of up to 25 days[26]. Therefore, these various drones can be used to establish the on-demand and multi-node quantum network connection for real-time coverage at different space and time scales, and the scale can be from a local-area network of hundreds of meters to a wide-area network up to hundreds of kilometers. By connecting such portable quantum nodes with the existing satellites and ground-based nodes, a quantum network with full and broad coverage can be expected, as shown in Fig. 1a.

Generally, the beam diffraction is a fundamental problem for a free-space quantum network, where diffraction loss dominates over large distances such as the satellite-to-ground distance. In a drone-based network, however, multi-nodes can be applied to divide a long link into shorter links. In this case, each drone node can receive a photon and retransmit it to the next node for cascaded transmission. Symmetric transmitter/receiver beam apertures and single-mode-fiber (SMF) coupling technology are keys for this multi-node quantum network architecture, and reasonable beam aperture can be used to achieve a satisfactory low link loss[27]. As shown in Fig. 1b, 2.79 dB loss can be achieved for a 100 km link with only 300 mm beam aperture, and such link distance is within the earth-curvature-limit at 20 km altitude, where a space-like scattering loss can be achieved in the turbulent-free air. In a smaller scale, a local network can be easily established with much smaller beam apertures and thus airframe weight/sizes (Fig. 1c and d), and a future portable quantum node can be as small as a modern commercial picture drone.

In this work, we demonstrate the first step towards such full-scale quantum network with a drone-based entanglement distribution. It is realized with symmetric transmitter and receiver beam apertures with SMF coupling technology, which makes it scalable for cascaded transmission. This entanglement distribution is achieved over 200 meters and a coverage duration of 40 minutes, with a Clauser-Horne-Shimony-Holt (CHSH) S-parameter of up to 2.49 ± 0.09. All-weather operations at daytime, clear night and rainy night are presented because of high signal-to-noise ratio in our system. Consequently, our drone node is highly reliable with all-day operation in even harsh environment. 12 dB node-to-node loss is achieved with a high-precision close-loop two-stage airborne acquiring, pointing, and tracking (APT) system, which can be further reduced with better optical alignment. The current quantum drone node is based on an octocopter with 35 kg take-off weight. Similar system with larger beam apertures can be adapted to high-altitude UAVs towards a multi-nodes network, and the broad area coverage can be expected.

In experiment, the main challenge for this drone-based entanglement distribution is to integrate the quantum node in a small drone. The octocopter we developed is with high thrust and low structure weight, and its payloads include an airborne entangled-photon source (AEPS) and two APT units. They are all home-built within a total weight of 11.8 kg including all the control electronics, which is the key to long flight duration.

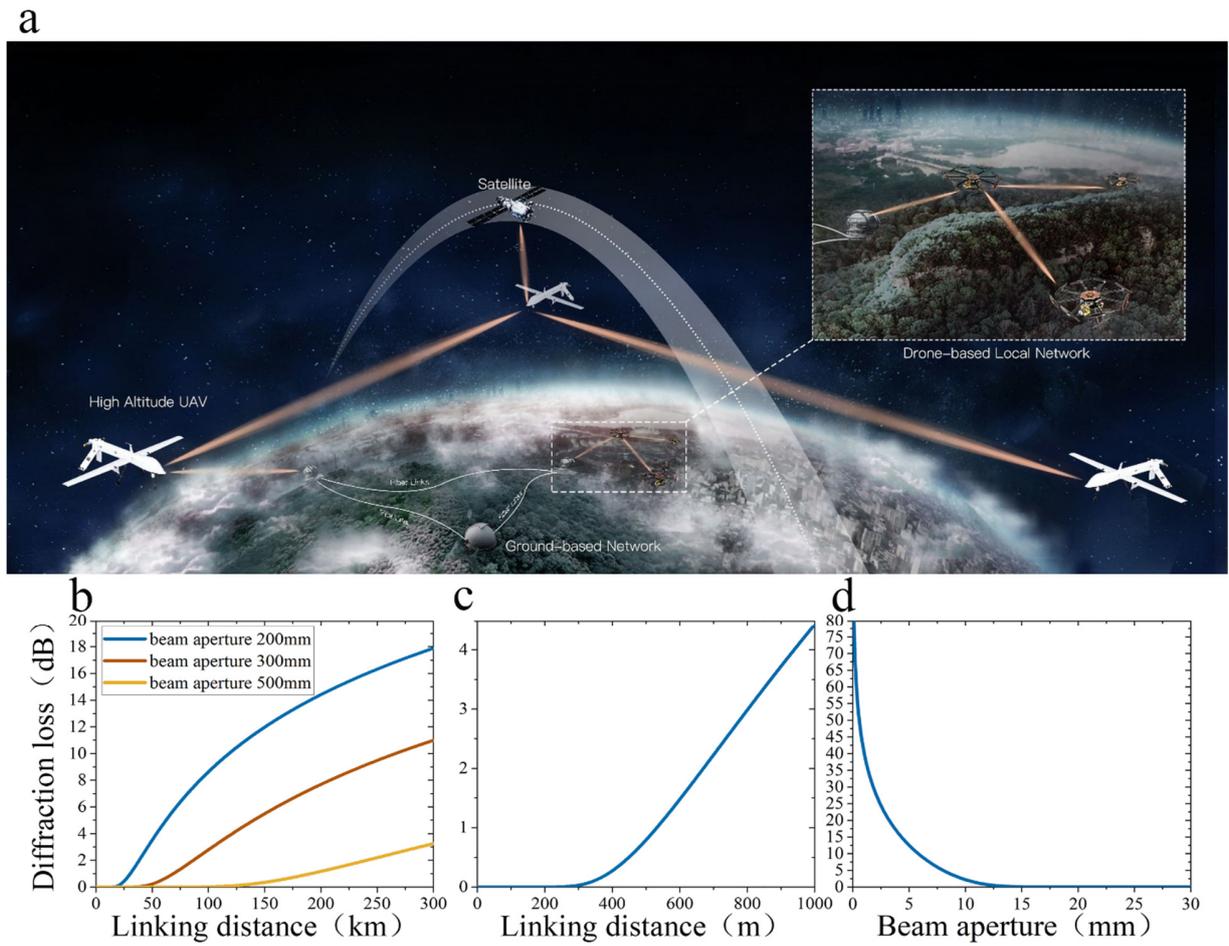

**Fig. 1 | Schematic of a scalable quantum network with drone-based link nodes. a,** Illustrative scheme for a drone-based quantum network from local to wide-area scales. The local-area network can be established with the plug-and-play drone nodes for fast connection with ground stations. While the wide-area network can be formed by high-altitude UAVs in cascade and connection with existing quantum satellites and ground fiber-based networks. **b,** Diffraction loss in a wide-area network. Up to 300 km link distances can be expected with reasonable loss and beam apertures using high-altitude UAVs. **c,** Diffraction loss in a local-area network with the beam aperture of 26.4 mm according to our experiment. **d,** Diffraction loss as a function of beam aperture at 100 m link distance according to our experiment.

The AEPS we build has a total weight of 468 g for an entangled photon pair generation rate of ~2.4 million per second under a pump of 15 mW. It is built with Sagnac interference of a two-sides pumped type-II spontaneous parametric down-conversion (SPDC)[28] from a periodically poled KTiOPO$_4$ (PPKTP) crystal, with schematic shown in Fig. 2a. By tuning PPKTP temperature, we achieve degenerate SPDC at 810.0 nm with 405.0 nm pump. When the half-wave plate HWP2 and the quarter-wave plate QWP are tuned to correct angles, the following maximum polarization-entangled photon state $|\psi\rangle = |HV\rangle_{12} - |VH\rangle_{12}$ can be generated, where $|H\rangle$ ($|V\rangle$) represents the horizontal (vertical) polarization state, and the subscripts 1 and 2 denote two output ports connecting to the Alice and Bob links, respectively. We performed four correlation measurements by projecting the photon in port 1 in $|H\rangle$, $|V\rangle$, $|D\rangle = (|H\rangle + |V\rangle)/\sqrt{2}$ and $|A\rangle = (|H\rangle - |V\rangle)/\sqrt{2}$ states, respectively, and recorded two-fold coincidence counts while changing the linear polarization projection angles in port 2. The visibilities are measured to be 97.4%, 97.7%, 97.1% and 97.0%, respectively, as shown in Fig. 2b. The CHSH Bell inequality was also tested, with the S-parameter measured to be 2.725 $\pm$ 0.017.

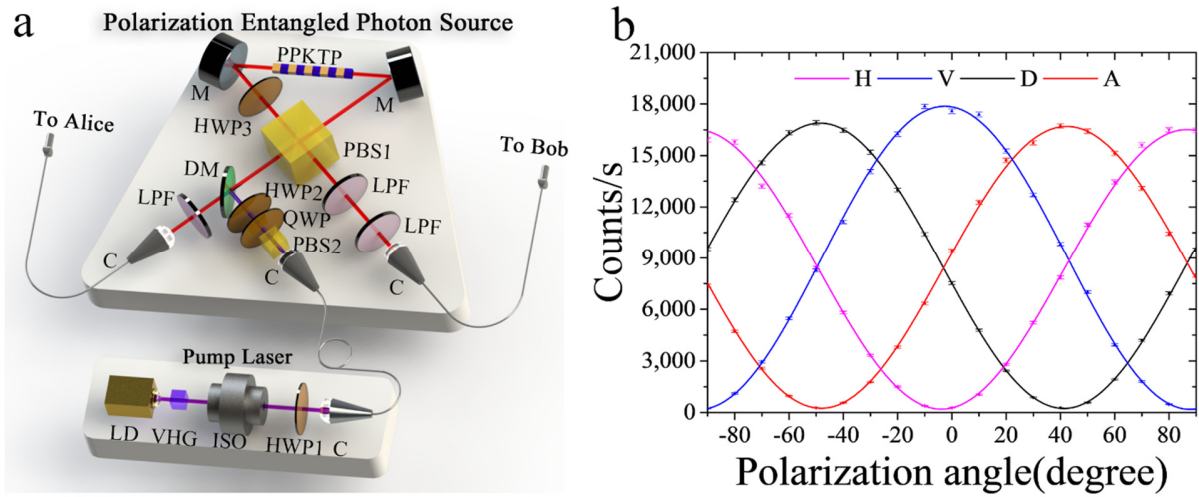

**Fig. 2 | Schematic of the AEPS and its performance. a,** Schematic map of AEPS. The pump laser is a 405 nm LD that is self-injection locked with a volume holography Bragg (VHG) grating. The polarization entanglement is generated in a Sagnac loop with a PPKTP crystal. LPF: long-pass filter, M: mirror, C: collimator. **b,** The measured two-photon correlation in lab. The interference with one photon projected to $|H\rangle$, $|V\rangle$, $|D\rangle$, and $|A\rangle$ states are all over 97.0 % visibilities. The error bars represent one standard deviation in the counts.

Another key to this successful entanglement distribution is the efficient air-to-ground optical links with SMF coupling. Such single-photon SMF coupling technology greatly enhances the signal-to-noise ratio for all-weather operation but adds more challenges for the tracking precision. Here it is achieved with a homemade APT system, including two pairs of airborne transmitter APT (TX) units for entanglement distribution and ground-based receiver APT (RX) units for Alice and Bob stations. These units are designed to realize bi-directional APT[29] for both upward and downward link directions so that both TX units and RX units can be correctly pointed, as shown in Fig. 3a. Symmetric beam size of 26.4 mm full width half maximum (FWHM) is used for both TX and RX units, which is sufficient for low-loss propagation of 810 nm entangled photons in free-space distance of over 100 m according to Fig. 1c. Each APT unit is composed of a telescope platform and a 3-axis motorized gimbal stage for two-stage tracking. The coarse tracking is achieved by the gimbal stage for pointing the whole telescope platform to a 940 nm laser diode (LD) target on the other link end. As shown in the insets of Fig. 3a, the fine tracking is achieved on the TX or RX telescope platform, where a position-sensitive detector (PSD) is used to lock a fast steering mirror (FSM) to the beacon light of 532 nm or 637 nm on TX or RX side. With proper feedback electronic controls, TX unit and RX unit can be pointed to each other within the accuracy for SMF coupling.

Before the quantum measurement, we performed a classical test of our APT system at a link distance of 100 m. The PSD tracking errors were recorded by onboard data logging system, and typical ones at TX unit and RX unit are shown in Fig. 3b and c. With the octocopter on the ground, the tracking errors are 1.19 μm for TX unit and 0.57 μm for RX unit, while they slightly increase to 1.33 μm for TX unit and 0.62 μm for RX unit after takeoff. All of them are smaller than the mode field diameter of 5 μm for the SMF at 810 nm. An 808 nm LD is used as the reference light and coupled in the transmitter fiber link with 1% power ratio through a 99:1 fiber coupler. The fiber-to-fiber coupling losses are measured to be 12 dB and 14 dB for Alice and Bob, respectively.

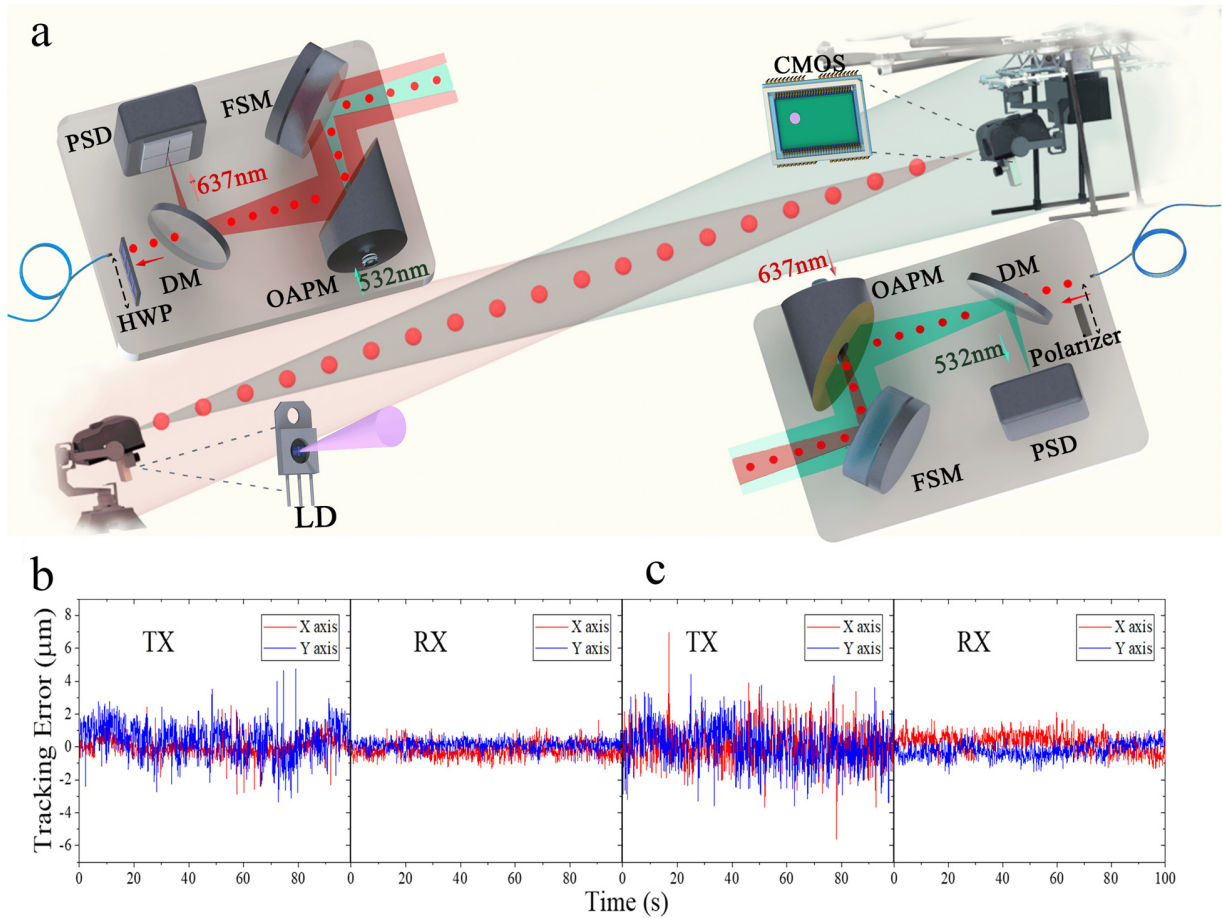

**Fig. 3 | Schematic of the APT system and its performance. a,** Schematic map of the whole APT system. A coaxial zoom camera is used on each APT unit to image the 940 nm LD on the other link side for coarse tracking. It generates the feedback signal to gimbal stage for moving the image to the center of CMOS array. The TX and RX telescopes are shown in details for the two-way find tracking using the 532 nm and 637 nm beacon light. **b,** Measured tracking errors of TX unit and RX unit before takeoff. **c,** Measured tracking errors of TX unit and RX unit during flight.

The total weight for each TX APT unit is 3750 g, which is within the payload limit of our octocopter. It cannot be lighter because of the bulky commercial parts used, which can be further reduced to fit onto a size of modern picture drone. The coupling loss is mainly attributed from the non-perfect alignment of the telescope and much stringent field overlap condition for SMF coupling, which can also be reduced in the future.

Between the AEPS and the TX APT units, we have included real-time polarization correction system to maintain the original entanglement state from the source against in-air vibrations from air turbulence. It offers full polarization control for real-time compensation, under the classical reference from an 808 nm onboard LD. Meanwhile, a classical communication link is utilized for coincident count synchronization from Bob to Alice through a classical optical fiber communication system.

Then our setup is ready for the entanglement distribution. Our test field limits Alice and Bob separation to 200 m, with each air-to-ground link of 100 m, respectively. We performed entanglement distribution in daytime, clear night and rainy night. The experimental scheme is shown in Fig. 4a, and we can make full projection measurement for the entangled photon pairs collected at Alice and Bob stations. The CHSH inequality[30] is used to characterize the entanglement, which is given by

$$S = E(a,\ b) - E(a,\ b') + E(a',\ b') + E(a',\ b') \tag{1}$$

where $a$ $(b)$ and $a'$ $(b')$ are projection angles at Alice (Bob), which are changed by the HWP sets inside the RX telescopes, and $E$ is the quantum correlation between the two stations, which is measured under the projection groups $(0, 1/8\pi)$, $(0, 3/8\pi)$, $(1/4\pi, 1/8\pi)$ and $(1/4\pi, 3/8\pi)$. The experimental results are shown in Fig. 4b. The CHSH S-parameters are calculated to be 2.41 $\pm$ 0.14, 2.41 $\pm$ 0.24 and 2.49 $\pm$ 0.09, in daytime with illuminance of 7316 lx, clear night and rainy night, respectively, without subtracting accidental counts. At all three above conditions CHSH-type Bell inequalities are violated with up to 5.4 standard deviations, which confirm successful entanglement distribution in our drone-based quantum network. Consequently, our drone node is highly reliable with all-day operation in even harsh environment. To the best of our knowledge, it is the first time to achieve this between a moveable node and ground ones.

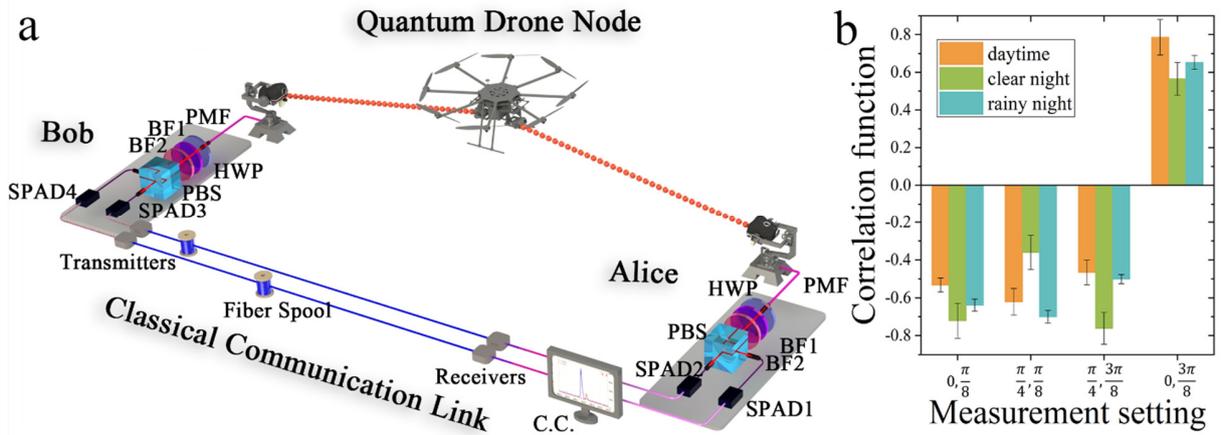

**Fig. 4 | Scheme and measurement results of the experiment. a,** Illustration of the experimental set up for the entanglement distribution experiment. The drone node distributes the entangled photon pairs to the Alice and Bob stations. A classical fiber communication link is used to send Bob photon clicks to Alice for coincidence measurements; BF1, BF2: bandpass filter. **b,** Measured correlation function results for the CHSH inequality calculation at daytime, clear night and rainy night. The error bars represent one standard deviation in the counts.

In this experiment, the octocopter is hovering during the current entanglement distribution. However, we have experienced a maximum wind speed of 18 km/h in the measurement, and the air turbulence can move the airframe around by up to ~1 meter. The APT system still offers good tracking for efficient SMF coupling and real-time polarization compensation system preserves the state for the Bell inequality violation. A faster polarization compensation scheme is under development to achieve entanglement distribution in high-g in-flight maneuver for future applications.

In conclusion, we have shown the first drone-based entanglement distribution for 200 meters distance. The CHSH S-parameter was measured up to 2.49 ± 0.09, and the Bell-inequality violations were obtained in daytime, clear night and rainy night. A light-weight polarization entanglement source and two high-precision airborne APT units have been integrated into an octocopter with 35 kg take-off weight. Symmetric transmitter and receiver beam apertures are used towards a scalable multi-node quantum network. The current result is sufficient for the realization of a drone-based local-area quantum network with 40 minutes on-demand coverage. With specially designed components, such local-area quantum network can be packaged to a much smaller size of a picture drone. This drone node can be scaled up with larger beam apertures and loaded on high-altitude UAVs for wide-area coverage. The link distance can exceed 300 km within the earth curvature limit and with reasonable aperture sizes, which can be further cascaded for larger scales. Such multi-node drone-based quantum network is essential to realize full coverage in multiple space and time scales, with the connection to the existing satellites and ground fiber-based networks.

**Acknowledgments**

The authors thank Wei Hu, Lijian Zhang, Xiaosong Ma for helpful discussions, Fullymax Co., Ltd. for developing the high energy density battery packs, TPpower Co., Ltd. for developing the high performance brushless motors, and Hobbywing Co., Ltd. for custom design the electric speed controllers. This work was supported by National Key R&D Program of China (No. 2017YFA0303700), National Natural Science Foundation of China (No. 51890861, No. 11674169, No. 11621091, and No. 91321312), and the Key R&D Program of Guangdong Province (Grant No. 2018B030329001).

**Author contributions**

Zhenda Xie and Yan-Xiao Gong conceived the original idea and designed the experiment. Hua-Ying Liu, Xiao-Hui Tian, Changsheng Gu, Pengfei Fan, Xin Ni, Ran Yang, Ji-Ning Zhang, Mingzhe Hu performed the measurements. All the authors helped on the manuscript preparation. Zhenda Xie, Yan-Xiao Gong, and Shi-Ning Zhu supervised the whole work.